**K-Prototype Segmentation Analysis on Large-scale Ridesourcing Trip Data**

**Jason Soria**
PhD Candidate
Northwestern University
Department of Civil and Environmental Engineering
Technological Institute, 2145 Sheridan Road, Evanston, IL 60208 USA
jason.soria@u.northwestern.edu

**Ying Chen**
Adjunct Professor
Northwestern University
Department of Civil and Environmental Engineering
Transportation Center, 600 Foster St., Evanston, IL 60208 USA
y-chen@northwestern.edu

**Amanda Stathopoulos**
Assistant Professor
Northwestern University
Department of Civil and Environmental Engineering
Technological Institute, 2145 Sheridan Road, Evanston, IL 60208 USA
a-stathopoulos@northwestern.edu

Word Count: 208 (abstract) + 5873 (text) + 1143 (references) + 250*(5 tables) = 8474

**ABSTRACT**

Shared mobility-on-demand services are expanding rapidly in cities around the world. As a prominent example, app-based ridesourcing is becoming an integral part of many urban transportation ecosystems. Despite the centrality, limited public availability of detailed temporal and spatial data on ridesourcing trips has limited research on how new services interact with traditional mobility options and how they impact travel in cities. Improving data-sharing agreements are opening unprecedented opportunities for research in this area. This study examines emerging patterns of mobility using recently released City of Chicago public ridesourcing data. The detailed spatio-temporal ridesourcing data are matched with weather, transit, and taxi data to gain a deeper understanding of ridesourcing's role in Chicago's mobility system. The goal is to investigate the systematic variations in patronage of ride-hailing. K-prototypes is utilized to detect user segments owing to its ability to accept mixed variable data types. An extension of the K-means algorithm, its output is a classification of the data into several clusters called prototypes. Six ridesourcing prototypes are identified and discussed based on significant differences in relation to adverse weather conditions, competition with alternative modes, location and timing of use, and tendency for ridesplitting. The paper discusses implications of the identified clusters related to affordability, equity and competition with transit.

## INTRODUCTION

Transportation Network Companies (TNCs) are prominent in many urban transportation ecosystems. The growth of ridesourcing patronage is attributed to the convenience compared to traditional modes. Widespread adoption of smartphones embedded with GPS technology has enabled travelers to street-hail rides through mobile applications, get real-time information about waiting times and make cashless e-payments as well as rating driver performance. More recently, major operators use algorithms to match passengers along similar routes in real-time, providing a new generation of shared rides. However, the proliferation of ridesourcing services has been a disruptive force in the mobility landscape and many questions have been raised about negative externalities and socio-spatial equity of supply. There is significant need for more insight on ridesourcing use patterns for cities to prepare policy, regulation and infrastructure plans. Yet, this mobility transformation has not been widely studied as many TNCs are reluctant to make their data publicly available. Recent data-sharing agreements with the City of Chicago, IL enables researchers to examine the role of ridesourcing in the transportation ecosystem using detailed temporal and spatial data on ridesourcing trips.

Several studies have characterized the adoption, frequency, and attitudes towards ridesourcing (1-4), but none have used trip data at the scale and scope provided by the City of Chicago. This study uses this newly released trip data, to develop insights about the role ridesourcing plays in the transportation ecosystem. The detailed spatio-temporal patronage data from operators Uber, Lyft and Via is merged with local transit, taxi and weather observations. The goal of this research is to investigate the variations in patronage of ride-hailing. By studying the emerging mobility patterns present in the data and examining the uneven locational and pricing patterns we develop insights about the motivations of users.

This paper utilizes an unsupervised learning algorithm to examine the underlying patterns in the data. Due to the mixed data types (i.e. data containing both numeric and qualitative/categorical variables), the K-Prototypes algorithm is selected (Huang (5). It is an extension of the K-Means algorithm that deals with categorical data. The model produces a classification of the data into K number of prototypes, similar to K-Means clusters.

This study contributes to the literature by providing a closer look at large, relatively disaggregate TNC data in a major metropolitan area. After tuning parameters for the best fit, the optimal number of prototypes to describe the data is 6. The first group of users (i.e. prototype) contains trips that occur in adverse weather conditions such as rainy weather. The second prototype involves trips that occur in the evening. The third prototype represents trips that are typically longer in distance but not shared. The fourth prototype is defined by trip origin and destinations being to the two major airports in Chicago: O'Hare and Midway. The fifth prototype is defined by short, solo trips occurring in areas that are well served by transit. The sixth prototype is defined by nearly all observations being shared rides.

This study is structured with the following sections. Following this section is a literature review that covers the state-of-the-art in TNC research. After the literature review is the methodology section that explains the model development and attribute selection. The next section reviews the algorithm's output and leads into the policy discussion. Finally, the conclusion contains a review of what is achieved in this study, its limitations, and possible future works.

## LITERATURE REVIEW

Since the inception of Uber in 2009, shortly followed by Lyft, ridehailing has already undergone significant service evolution as discussed by Shaheen and Cohen (6). Among these, shared ride-hailing, or ride-splitting, matches individuals in real time based on shared routes, and microtransit, or curb-to-curb options, match users into van-sized vehicles based on dynamic or planned routing. For the major TNCs, riders can now typically decide between solo travel or rode-splitting in the same application. Authorizing sharing typically results in lower fares but longer travel times as the trip now includes several stops that may cause the vehicle to deviate from the optimal path for a single origin-destination pair.

Researchers have tried to develop a better understanding of ridesourcing travel, but data is scarce. Uber and Lyft do not publicly share their data, or do so with significant aggregation of rides, causing a dearth of empirical studies. Owing to empirical data scarcity, researchers have gotten creative to gain insights on usage and possible impacts. Henao and Marshall (7) went so far as to become a TNC driver to collect trip information. The research suggested extra vehicle miles were generated by deadheading, a form of inefficiency that is difficult to map using stated data. After accounting for deadheading mileage, the average occupancy of a TNC vehicle is less than one person. Such empirical approaches can offer new insights but some caveats limit the generality. The trips recorded in this study are a relatively small sample of total rides in the entire region and are biased due to all data coming from a single driver. Other groups have utilized their own ridehailing experiments to further highlight competition with transit (8) and equity compared to taxis (8, 9) further highlighting the need for a comprehensive dataset for analysts to access. Along with access to data, connections between stated and revealed preference data may also lead to more informed policy decisions (10).

The current understanding of ridesourcing travel is mostly informed by survey research. In the following we briefly overview relevant ridesourcing work and relate findings to the current analysis of real large-scale data. Several studies delve into the trip purposes of ridesourcing trips. Defined by its utilization of large capacity vehicles, micro-transit (also known as demand-responsive transit or on-demand transit) can serve as a tool to address public transit overcrowding and the first-last mile problem (11). It is mainly utilized to commute (11, 12). Instead, trips made by the more taxi-like TNCs are mostly for social/recreational trips (7, 13-15). Trip purpose is not included in the current analysis due to the data anonymization. However, in future works spatial examination of locations of interest combined with other trip attributes can be used to infer trip types.

The effects of TNCs on the transportation system, specifically via the competition or complementarity on other modes, is a core area of research. In particular, due to the similarity of the services, the impact on taxis has been widely studied. TNCs have significantly reduced the demand for traditional taxi services such that taxi drivers altered their strategies to remain profitable (16-21). Focusing on transit competition, Schwieterman and Smith (8) find that TNCs are preferred over public transit, especially when origin-destination pairs are not well served by transit. Along similar lines, ridesourcing's relationship with public transit is found to be complementary within large cities with small transit agencies (22). Further determinants of ridesourcing use relate to the travel environment. Frei, Hyland (23) found that weather affects TNC usage. Though their study focused on micro-transit, other TNC services may also be affected by adverse weather.

Owing to these observations, the ridesourcing trip data used in this project are matched in space and time with data on weather, equivalent origin-destination transit travel times, and peak taxi demand.

**MODEL DEVELOPMENT**

The data analysis used to examine patterns of ridesourcing use in this project is an unsupervised learning technique called K-Prototypes. K-prototypes is similar to K-means since both aim to cluster several observations together according to their attributes. The advantage K-prototypes has in this situation is its ability to also accept categorical variables. More details on K-prototypes development can be found in Huang (5).

The challenge of dealing with categorical variables has been considered for segmentation analysis. The problem is that the K-means algorithm relies on all variables to be numerical. Specifically, in the K-means algorithm for a continuous variable such as travel time, the distance between an observation's travel time and the proposed cluster's mean travel time is the key element for identifying clusters among observations. With a categorical variable such as vehicle type, the distance is no longer applicable. One strategy to include categorical variables in the K-means algorithm is to code each category as a dummy variable (0 or 1). The distance calculated by K-means algorithm for a categorical variable is then 0 or 1 which is not informative. With the K-prototypes algorithm the mode of the category is used and a measure of a matching coefficient is used. The formulation from Huang (5) of K-prototypes algorithm is summarized in equations 1 to 4.

The matching of observations to prototypes involves reducing the error or cost function. This cost function represents the distance between observation data and the assigned prototype center. **Equation 1** shows that the error, $E$, is the sum of distances from the prototype center. $X_i$ are the attributes of trip $i$, $Q_l$ is the center of prototype l, and $y_{il}$ is a dummy variable that is equal to 0 when trip $i$ is assigned to prototype l. It is then the sum of squared distances for n TNC trips across k number of prototypes. **Equation 2** breaks down $d(X_i, Q_l)$ into numerical and categorical components, where the first term is the squared numerical distance of attribute $j$ of trip $i$ from the center for attribute $j$ of prototype l; the second term includes a component to determine the weight, $\gamma_l$, of the categorical variables to the total error $E$. The error of prototype l is then calculated in **Equation 3**, where $E_l^c$ is further explained by **Equation 4**. $C_j$ is the set of all unique values of categorical attribute $j$, and $p(c_j \in C_j \,|l)$ is then the probability of unique value $q_j$ from set $C_j$ being in prototype $l$.

$$E = \sum_{l=1}^{k} \sum_{i=1}^{n} y_{il} d(X_i, Q_l) \tag{1}$$

$$d(X_i, Q_l) = \sum_{j=1}^{m_r} \left(x_{ij}^r - q_{lj}^r\right)^2 + \gamma_l \sum_{j=1}^{m_c} \delta\left(x_{ij}^c, q_{lj}^c\right) \tag{2}$$

$$E_l = \sum_{i=1}^{n} y_{il} \sum_{j=1}^{m_r} \left(x_{ij}^r - q_{lj}^r\right)^2 + \gamma_l \sum_{i=1}^{n} y_{il} \sum_{j=1}^{m_c} \delta\left(x_{ij}^c, q_{lj}^c\right) = E_l^r + E_l^c \tag{3}$$

$$E_l^c = \gamma_l \sum_{j=1}^{m_c} n_l (1 - p(q_{lj}^c \in C_j | l)) \tag{4}$$

The advantage of using K-Prototypes algorithm over other clustering algorithms is highlighted by **Equation 4**. A common way to code categorical variables for other data-driven methods is to use one-hot encoding. Using this method, the unique values of a category are coded as a dummy variable equal to 1 when denoting the variable of interest and 0 otherwise. Algorithms using one-hot encoded data fail to recognize that these unique values belong to a categorical variable because categories are reduced to 0 or 1. The advantage of the K-Prototype algorithm lies in the recognition that these values denote categorical variables and using the probability of a unique value from a set $C_j$ being in prototype $l$.

This model is implemented and tuned with the R programming language using the 'clustMixType' package (24, 25). Using this package, the error is minimized and the weighting of the categorical error is optimized. Much like other clustering methods, the number of prototypes is a tunable parameter. The final tunable parameters are discussed in the results section.

## DATA DESCRIPTION

The data used in this project is drawn from TNC trip data provided by the City of Chicago (26). The trip data begins on November 1, 2018 and is updated monthly. For the purpose of obtaining lower optimization times and being able to match the equivalent transit travel times, the data is partitioned to weekdays in November 2018. Holidays are not included. This leaves a total of 3,085,070 trips in the dataset. The trips are grouped at the census tract level and include variables such as travel time, travel distance, fare, whether it was a shared trip (and if it was how many other passengers were included), census tract origin-destination pairs, and timestamp of pickup and drop-off rounded to the nearest 15-minute increment.

The weather data were collected from OpenWeatherMap specifically for the City of Chicago in November 2018 (27). The data is at the hourly level and includes amount of rain and snow in the previous hour, qualitative description of the weather (such as raining, hazy, sunny, etc.), and temperature. The station collecting the data is located at O'Hare International Airport at the northwest tip of the city limits. The supplementary transit travel times dataset was created for each unique origin-destination-time-day tuple. Transit travel time estimates were obtained using the Google Distance Matrix (Advanced) API by providing the census tract of origin, the census tract of destination, travel mode (transit), and departure time (28). From the API, approximate transit travel times between point to point origin-destination pairs are collected. This data include the expected access (walking to transit stop) and expected wait time just as one would view from a navigation assistant device/app. Since the data were only available from 6AM to 10PM for much of the network, the TNC trips data which is publicly available and is collected by the City of Chicago, was also restricted to these hours. Even when available, transit is typically operating with reduced capacity after 10PM, so to enable a fair comparison we restrict the analysis to these regular travel hours. The taxi trip data used in this research correspond to peak taxi demand in 2014 and are further described by Chen et al. (29). The data are the monthly taxi trips between census tract origin-destination pairs, which are referred to as Monthly Taxi Frequency later in the analysis. This data are included to characterize and compare the spatial relationship of taxi usage by matching each ridesourcing trip with the total taxi flow between the same origin and destination. Table 1 contains descriptive statistics for the numerical analysis data.

## ANALYSIS OF RESULTS

During the estimation phase, the K-prototype algorithm was tuned to select the optimal number of prototypes. This was determined by developing models including a number of prototypes ranging from 2 to 14 and calculating the total cost across all observations. The final number of prototypes chosen is 6 based on interpretability of segmentation variables and guidance from the plot which in figure 1 shows a clear "elbow" at 6 prototypes (30). An elbow occurs when adding more clusters does not sufficiently improve the objective function. $\gamma$ is the tradeoff between numerical cost and categorical cost optimized by the 'kproto' function in the 'clustMixType' package and is estimated to be 1.33 for all prototypes as per **Equation 2** and **4** (25). There is no intuitive meaning to this value except that it can be user-specified, and higher values mean that the categorical variables receive a higher weight. **Figure 1** shows how many observations belong in each prototype cluster. A summary of the top 6 origin and destinations, respectively, are given in **Table 2**. The clustering results are shown in **Table 3** along with mean values of explanatory attributes in each prototype.

**TABLE 1 Descriptive Statistics of Ridehailing, Transit, Taxi and Weather data**

| Numerical Variable | Median | Mean (Standard Deviation) |
|---|---|---|
| Travel Time (minutes) | 13.32 | 15.47 (9.98) |
| Distance (miles) | 2.70 | 3.79 (3.19) |
| Total Fare ($) | 10.00 | 11.24 (6.27) |
| Parties Joined in Trip | 1 | 1.32 (0.77) |
| Humidity (%) | 71.00 | 73.58 (11.89) |
| Wind Speed (mph) | 3.00 | 3.82 (2.33) |
| Rain last hour (inches) | 0.00 | 0.061 (0.26) |
| Minute after Midnight | 930.00 | 887.5 (268.20) |
| Transit Travel Time (min) | 17.95 | 21.10 (15.40) |
| Monthly Taxi Frequency | 1004 | 14,976 (36,875.57) |

An important observation related to variable selection in the presence of potential correlation needs to be made. In practice, transportation modelling often deals with concerns surrounding the correlation among time, distance and cost, either by interacting or dropping variables. Yet, ridesourcing represents a special case due to the dynamic demand-responsive pricing that relaxes this typical correlation. While we cannot separate out instances of surge-pricing from this data we note that some interesting relationships are discovered when comparing prototypes. Notably, while the variables are correlated, *on average*, within the specific clusters the relationship reveal vast differences in per mile costs. Table 4 and related discussion highlight these insights.

We now turn to summarize the contours of the six user clusters. On the whole, the analysis did not produce prototypes that were heavily differentiated by temperature or snow fall in the past hour. Yet weather effects were evident in the first segment of users (Prototype 1 or P1_weather). P1_weather is the second largest prototype and is characterized by its relatively low total fares and short travel times and distances. This short-distance travel, averaging 4 miles, is coupled with the strongest weather impacts observed, namely the presence of adverse weather seen with rain, humidity, and wind speed. The distinct nature of prototype 1 suggests the use of ridesourcing for short distance travel to cope with adverse weather in the early part of the day.

Prototype 2 (P2_late-night) is the largest segment with 30.2 percent of users. While still representing shorter trips, it is distinct from P1 due to the trip timing in the evening (average is 1080 minutes after midnight or 6PM) and the lack of relationship to weather conditions. Observing Table 5, these trips are most heavily focused in the wealthy downtown and near north areas. Furthermore, Table 2 illustrates that trips in this cluster originate from areas with the highest bar and tavern densities. This sizeable cluster suggests a strong tendency to use ridesourcing for evening travel which is in line with findings from Lavieri and Bhat (31).

**TABLE 2 Community Area Characteristics**

| Community Area | Per Capita Income ($) | Bar and Tavern Density (per sq. mi) | Transit Access Time* (min) |
|---|---|---|---|
| Chicago Average | 32,534 | 4.78 | 19.75 |
| Near North Side | 91,948 | 32.66 | 13.00 |
| Near West Side | 50,394 | 10.51 | 10.57 |
| West Town | 54,429 | 11.86 | 11.03 |
| Loop | 77,722 | 46.84 | 9.53 |
| Lincoln Park | 73,965 | 13.43 | 12.94 |
| Lake View | 67,066 | 19.15 | 11.17 |
| Midway | 28,925 | 3.27 | 33.79 |
| O'Hare | 27,212 | 0.17 | 84.64 |

Prototype 3 (P3_solo-non-transit) has longer travel times which tend to be associated with longer distances (albeit not associated with airport travel) and higher total charges. This large user segment (20.4 % of usage) suggests some transit gap-filling capacity of ridesourcing in Chicago whereas the origin-destination and time-matched potentially available transit trip would take 30% longer on average with transit travel-time taken as base. Notably, considering the fixed transit pricing of $2.25, the ridesourcing trips were on average six times more costly. Trips in this prototype are also typically not shared and concentrated in wealthier areas. This finding mirrors observations by Schwieterman and Smith (8) that ridesourcing is used even in areas with a wealth of transit options, although our analysis suggests that transit speeds are relatively low (Table 4) a factor that is easily tracked by travelers using real-time smartphone navigation tools.

Prototype 4 (P4_airport) represents a small group of users with long travel times dominated by trips to and from the main airports O'Hare or Midway International Airports (**Table 5**). This prototype also has trips where the origins and destinations are not served well by transit as seen with the average transit travel time being more than 70% longer. Along with poor transit connectivity, this cluster features relatively low taxi frequency. The low taxi frequency shows low demand for taxis between similar airport-based trips, likely because airport trips are relatively infrequent and can be completed by carpooling with known associates such as a family member or friend. These trips' fares are more expensive than in other prototypes, but relative to the cost of traditional taxis, are still affordable. Given that it is also more convenient to utilize ridehailing than to ask a family member to drive, the strong connection between airport travel and ridehailing is unsurprising. Taxi pickups at airports are declining and other revenue streams such as parking and rental cars are also negatively impacted (32, 33). This prototype highlights the strong competitive position against both transit and taxi for airport access, albeit it does not account for the issue of waiting time that might change this assessment in particular considering departures from Chicago airport where TNCs have limited access.

Interestingly, prototype 5 (P5_transit-competitive) is a small cluster that stands out as representing the shortest trips and for being the only case where trips could have been served better by transit. Notably, the average transit travel times would be 12.44% lower than the observed TNC travel times. This is a stark contrast with other prototypes as table 4 shows that most other prototypes' transit travel times are at least 30% longer than the ridesourcing equivalent ride. Most of these trips are in the Chicago Loop or just north of it where transit is highly concentrated in the core commercial area.

**TABLE 3 Prototype Attribute Results and Percentiles**

| Prototype | Travel Time (min) | Distance (miles) | Total Fare ($) | Parties Joining Trip** | Humidity (%) | Wind Speed (mph)** | Rain last hour (inches)** | Minute after Midnight | Transit Travel Time (min) | Monthly Taxi Frequency | Percent ridesplitting (%) |
|---|---|---|---|---|---|---|---|---|---|---|---|
| P1_weather (Percentile) | 637.40* (36th) | 2.16 (40th) | 9.05 (41th) | 1.07 | **82.96 (79th)** | **4.32** | **0.15** | 702.5 (37th) | 844 (39th) | 11695 (76th) | 18.77 |
| P2_late-night | 600.9 (33rd) | 2.07 (38th) | 8.85 (41st) | 1.07 | 66.07 (31st) | 3.57 | 0.01 | **1080 (72nd)** | 804 (37th) | 10031 (75th) | 17.65 |
| P3_solo-non-transit | 1284.0 (78th) | 5.74 (80th) | 15.56 (85th) | 1.04 | 72.29 (54th) | 3.66 | 0.03 | 878 (45th) | **1838 (78th)** | **3558 (64th)** | **10.48** |
| P4_airport | **2014.0 (94th)** | **12.25 (97th)** | 27.64 (97th) | 1.17 | 75.09 (61st) | 3.96 | 0.08 | 826.4 (40th) | **3392 (97th)** | **3840 (65th)** | 16.58 |
| P5_transit-competitive | 572.2 (31st) | 1.39 (21st) | 8.40 (40th) | 1.12 | 73.64 (56th) | 3.70 | 0.05 | 815.9 (39th) | **501 (19th)** | **144687 (98th)** | 13.76 |
| P6_ridesplitting | 1320.0 (79th) | 4.83 (74th) | 7.43 (15th) | **3** | 74.12 (59th) | 3.66 | 0.05 | 870.7 (45th) | **1545 (69th)** | **5286 (68th)** | 100 |

* **Bold** type indicates important feature
** Non-continuous variable with low range do not have percentiles included

**Table 4 Prototype specific Average Costs and Speed**

| Prototype | Average $ per Mile Traveled | Average $ per Minute Travel Time | Average Speed (mph) | % transit travel time above ridesourcing equivalent trip* |
|---|---|---|---|---|
| All Trips | 2.97 | 0.73 | 12.16 | 36.39 |
| P1_weather | 4.19 | 0.85 | 12.20 | 32.41 |
| P2_late-night | 4.28 | 0.88 | 12.40 | 33.80 |
| P3_solo-non-transit | 2.71 | 0.73 | 16.09 | 43.15 |
| P4_airport | 2.26 | 0.82 | 21.90 | 68.42 |
| P5_transit_competitive | 6.04 | 0.88 | 8.75 | -12.44 |
| P6_ridesplitting | 1.54 | 0.34 | 13.17 | 17.05 |

* = (Transit Travel Time – Ridesourcing Travel Time) / Ridesourcing Travel Time

**Table 5 Prominent Prototype Origins and Destinations**

| Prototype | ORIGINS | | DESTINATIONS | |
| --- | --- | --- | --- | --- |
| | Community | % in Prototype | Prototype | % in Prototype |
| P1_weather | Near North Side | 22.62 | Near North Side | 24.22 |
| | Near West Side | 13.63 | Loop | 15.40 |
| | West Town | 9.638 | Near West Side | 14.15 |
| | Loop | 9.537 | West Town | 5.280 |
| | Lincoln Park | 5.878 | Lincoln Park | 5.012 |
| | Lake View | 5.169 | Lake View | 4.561 |
| P2_late-night | Near North Side | 24.96 | Near North Side | 23.78 |
| | Near West Side | 12.84 | Near West Side | 13.16 |
| | Loop | 12.11 | West Town | 8.932 |
| | West Town | 7.502 | Lincoln Park | 8.162 |
| | Lincoln Park | 7.224 | Loop | 8.085 |
| | Lake View | 7.103 | Lake View | 7.664 |
| P3_solo-non-transit | Near North Side | 16.77 | Loop | 18.21 |
| | Loop | 10.47 | Near North Side | 12.07 |
| | Lake View | 9.795 | Near West Side | 11.25 |
| | Near West Side | 8.263 | Lake View | 7.916 |
| | Lincoln Park | 7.144 | West Town | 4.967 |
| | West Town | 6.115 | Lincoln Park | 4.723 |
| P4_airport | **Midway\*** | **13.80** | **O'Hare** | **16.17** |
| | **O'Hare** | **9.523** | **Midway** | **15.08** |
| | Near North Side | 7.606 | Near North Side | 9.966 |
| | Loop | 6.306 | Loop | 7.152 |
| | Near West Side | 5.624 | Near West Side | 6.974 |
| | Lake View | 4.607 | Lake View | 3.531 |
| P5_transit-competitive | **Loop** | **45.57** | **Loop** | **54.35** |
| | **Near North Side** | **32.15** | **Near North Side** | **21.88** |
| | Near West Side | 8.719 | Near West Side | 8.013 |
| | Lake View | 5.986 | Lake View | 5.982 |
| | West Town | 3.156 | West Town | 4.160 |
| | Lincoln Park | 2.688 | Lincoln Park | 3.112 |
| P6_ridesplitting | Near West Side | 13.81 | Near North Side | 14.90 |
| | Near North Side | 11.54 | Near West Side | 13.20 |
| | Loop | 10.60 | Loop | 13.18 |
| | West Town | 7.686 | West Town | 6.060 |
| | Lake View | 6.510 | Lake View | 6.058 |
| | Lincoln Park | 5.489 | Lincoln Park | 4.873 |

**\*Bold** type denotes important prototype features

Prototype 6 (P6_ridesplitting) with 12.8% of users is defined by representing nearly all shared authorized trips. This segment appears to reflect a more cost-conscious user group given that the ridesourcing price per mile is the lowest, and the competition in terms of price and time is closer to the potentially available transit trip.

To further understand motivations of different users Table 4 highlights the insights from comparing trade-offs within clusters, namely fare per mile, fare per minute, and average speed to the average reference of all ridesourcing trips. Table 4 shows that P1_weather, P2_late-night, and P5_transit-competitive prototypes have a more premium fare point with higher fare per mile and fare per minute than its counterparts. The results also show steep discounts for P6_ridesplitting as it has the lowest fare per mile and fare per minute. These results confirm the prototype interpretations as premiums are expected (through surge pricing or similar dynamics) for rides in bad weather, late at night when drivers may be few and potential-riders are unable to drive due to inebriation, and transit-competitive trips mostly occurring in the Loop community area which is the core commercial area. Discounts are also expected to appear with the ridesplitting prototype as reduced fares are expected with delays incurred by the detours when picking up a different party.

## DISCUSSION

The K-prototypes analysis is geared at finding relationships in the ridesourcing data by grouping similar observations together. The merging of multiple datasets further enables the prototypes search to identify the main ridesourcing profiles with regards to trip attributes (e.g. travel time, fare, origin and destinations, being private or shared), and competing mobility services (transit and taxi) along with weather conditions. This discussion section focuses on how the results relate to current research and can inform future research directions. Four areas of investigation are highlighted, centering on weather impacts, competition with transit and taxi, ridesplitting patterns and spatial distribution of ridesourcing.

### *Weather-dependence*

We find that while weather does not have a pervasive impact on ridesourcing across clusters, it does strongly determine the choices in P1_weather highlighted by its higher average windspeed, humidity, and rainfall in the last hour. The identification of this prototype gives evidence that weather can have a significant impact on TNC usage for as many as 25 % of trips. Taken together with results from Frei, Hyland (23) demonstrating weather impacts in a micro-transit choice experiments, this illustrates the importance of including weather as an explanatory variable in future TNC analyses. Inclusion of weather-variables in TNC analyses can further explain the interactions between ridesourcing and other modes. For example, weather was shown to impact active modes of transport, so including weather as an explanatory variable between the relationship of ridesourcing and active mobility can inform their demand in the future (34). This is especially useful for understanding how TNCs might relate to bikeshare as adverse weather has been shown to decrease its demand and contribute to increased ridership of other modes (35). find that ridesourcing demand increases during adverse weather conditions and compared the supply of TNC drivers to taxis. Their results illustrate the benefit of TNCs – in particular its dynamic pricing – over taxis as a tool to increase the supply of drivers and meet consumer demand.

### *Mode-substitution with transit and taxi*

The importance of understanding the relationship TNCs have with other modes is further highlighted by P4_airport and P5_transit-competitive prototypes. The airport prototype shows that airport trips are a major source of demand for ridesourcing because it provides more effective service than current transit options for many users.

The transit-competitive prototype illustrates the competitive nature beyond travel time of TNCs. Though Figure 2 shows that this is a smaller portion of the trips, representing only 5.1% of the data, this is still an interesting prototype because it emphasizes how TNCs offer several advantages that go beyond shorter travel times. As discussed by Lavieri and Bhat (31), this is troubling because ridesourcing's relationship with transit is complex as solo rides do not necessarily substitute transit trips. With shorter transit travel times and some demand previously met effectively by taxis, there is a need to map out the difficult to measure variables such as comfort, safety, and convenience that must be considered in conjunction with travel time. These insights may be critical to understanding the differing user perspective towards solo and shared ridehailing.

The relationship between taxis and ridehailing is more straightforward as the services are more comparable. While the literature review section briefly discussed changes in the taxi industry, a thorough investigation of the interaction between these modes is completed by Nie (16). Ridesourcing is an attractive alternative to taxis, however, there still remains a role for taxis in the transportation system as they remain competitive in highly dense areas during peak commuting hours. The substitution of taxis for ridesourcing also (though unintended) led to improved mobility equity in struggling communities as it is an option for those who do not possess bank accounts, credit cards, or smartphones (37).

### *Ridesplitting patterns*

Another major area of the literature is on the potential for TNCs to be a more efficient people mover than privately driven vehicles. The dynamic ridesharing literature examines the efficiency gains of ridesplitting over private modes (38, 39). Despite theoretical findings on the advantages of ridesplitting, there has been limited exploration of how this functions in real systems. A notable result from this work is the low share of split rides despite a relatively high share of riders indicating that they are willing to share their ride. For the complete dataset, 26.7% of all trips were authorized to be shared but of these only 68.5% were actually shared. That implies that only 18.3% of the overall rides were truly pooled, likely reflecting a lack of matching travel itineraries that were close enough in space and time for the matching to occur. The percentage of authorized shared trips of all prototypes except for P6_ridesplitting is well below the 26.7% figure.

When compared to the other prototypes, the ridesplitting prototype shows that pooled trip making can be seen as a separate profile of use. To further examine the patterns of ridesplitting, figure 3 shows the number of trips by separate trip-makers within a pooled trip for each prototype. The ridesplitting prototype has a much higher share of pooled trips including more than 3 riders. However, this prototype only constitutes 12.8% of the data. With such a small share of trips being shared, decision-makers that support TNCs should consider strategies that increase the number of pooled trips.

### Spatial patterns of use

Lastly, we discuss the spatial distribution of travel. Notably, the majority of trips occur in or around the Chicago Loop or airports with standouts Near North Side and Near West Side where there are typically more residential units than in the Loop and overall higher density compared to the rest of the city. Table 5 confirms that the top 6 origins and destinations hardly differ across prototypes. The strong concentration of flows is further illustrated in Figure 4 that shows the location of the top O-D pairs distinguished by bold borders. These areas tend to have higher influx of visitors, along with more leisure landmarks such as restaurants and night clubs. The residents of these community areas tend to have higher average incomes and possess higher educational attainments than the average Chicagoan. These results are in line with findings from Clewlow and Mishra (40) as those who are college-educated, younger, and living in denser areas are more likely to adopt ridehailing.

### POLICY IMPLICATIONS

This study identified several patterns of ridehailing usage across Chicago that highlight the need for careful policy implementation. The discussion of policy implications will focus on modal interactions and ridesplitting due to the need for insights to guide ongoing efforts to tweak fares, promote partnerships and regulate ridehailing to better serve the comprehensive mobility needs of Chicago residents. The core question that need to be explored relate to a) the challenge to provide effective service in areas with poor (or strong) transit options and b) to promote equity in hailing-access by understanding and promoting more affordable ride-splitting. Because ridehailing has been a disruptive innovation and lack of access to a comprehensive dataset on TNC activity, there is limited understanding regarding its relationship with other transport modes and the variation in ride-splitting adoption.

Much of the policy debate has focused on determining whether ridehailing is complementary or a substitute to other modes; this section discusses strategies that may facilitate synergy in the transport ecosystem.

### Ridesourcing and air-mobility accessibility

Given the identification of an airport prototype with strong connections to the core commercial areas of Chicago, one major policy trend has been to control ridesourcing's effect on airport infrastructure. Examples of this include extra fees to ride into airports and curbside management of drop-offs and pickups. This prototype serves as evidence for continued development of policies that will better manage the relationship between airports and urban mobility including prominent use of ridesourcing. With this prototype showing a strong connection between the commercial core of Chicago, policies should focus on connections that will appeal to business travelers. This remains a challenging area of research as new options including Vertical Urban Air Mobility is being tested in initiatives such as UberElevate with electric vertical takeoff and landing vehicles (41, 42). This highlights the need to craft regulations and partnership arrangements such as security checkpoints and luggage drop-offs (43). The rise of new services also highlights renewed equity and affordability concerns as they might give rise to further erosion of transit options.

### Ridesourcing and transit performance

Conversely, the airport prototype also suggests the need for policies to improve transit connections between downtown Chicago and the airports. The segmentation analysis revealed some intriguing patterns of competition. Ridesourcing appears to be used by a small group of users even when transit is seemingly the better option (P5_transit-competitive 5.1%) and at the same time, a sizeable segment will turn to their mobility-apps in areas where transit is in abundant supply but time-performances is poor (P3_solo-non-transit 20.3%). This opens a debate about perception and motivations of users, communicating options to travelers and developing new partnerships.

With the transit-competitiveness prototype showing that there are real possibilities for transit to be faster than ridesourcing, a practical policy effort is to improve the dissemination of transit information. Local transit agencies can develop Advanced Traveler Information Systems that highlight cases where transit is competitive to increase their ridership (44). Other strategies could be used in conjunction with MaaS in multi-modal systems to nudge riders towards transit. Studies have shown that travel behavior can be influenced using soft strategies (45). These strategies such as making transit the default option or highlighting the broader benefits of supporting transit through patronage can be facilitated through navigation application. While this type of policy improves transit-competitiveness, ridesourcing may still be dominant in many areas and promotion of sharing is vital in this situation.

### More ridesplitting?

Promoting ridesourcing naively may worsen traffic conditions, however, promoting shared rides to increase the demand for ridesplitting may be a reasonable solution. Policies that incentivize shared rides such as a tax that increases fees for exclusive rides could lead to higher demand for sharing and increased transit ridership (46). The trade-off between delays and lower fares could be used to promote sharing and even increase mobility for disadvantaged groups where high fares turn them away. Policies providing travel support for unemployed and low-income residents via vouchers or further lowering fares increases travel and opportunities when other modes are not feasible. The ongoing debate in Chicago and cities around the US has focused on the lack of broader coverage, outside transit rich areas, of ridesourcing. Figure 4 highlights the lower share of rides occurring in and between historically underserved communities on the South and West sides of Chicago. Policies geared at promoting shared ridesourcing between underserved areas is an opportunity to reduce Vehicle Miles Traveled and to support disadvantaged communities.

### CONCLUSION AND FUTURE WORK

This study examines a unique TNC dataset from Chicago, IL by utilizing the unsupervised learning K-prototypes algorithm that accepts categorical data. The goal of this study is to identify patterns of TNC patronage regarding service attributes, weather, transit, taxis, characteristics of origins and destinations, and ridesplitting. The analysis revealed 6 distinct ridehailing user segments. The segments were identified in relation to adverse weather conditions, evening trips, longer trips, trips to the airport, trips that would be better served by transit, or trips that are pooled. The segments are discussed in the context of relative performance of ridesourcing as well as examining the origin and destination of flows to better interpret the spatial and performance variation.

The identification of these distinct trip types shows where future research is warranted. The discussion in this study focuses on how future research should consider factors such as weather and other external factors when estimating the demand for TNCs and other modes, airport-based mobility options in the future, understanding why TNCs have competitive advantages besides faster travel times, and why more trips are not shared. The last point made in the discussion emphasizes how most of the trips are completed in and surrounding the CBD of Chicago. In summary, the concentration of trips in the downtown area where of mobility options and amenities are abundant, along with notable variation in performance of ridehailing across user clusters, prompt a deeper discussion of *where and for whom* ridehailing enables mobility.

The main limitations of this study come from the constraints of the merged data-sets. Firstly, the weather data is collected at only one location. Considering the size of Chicago and the location of the station, the data may not be representative of local weather. Secondly, the TNC, taxi, and transit data are aggregated at the Census tract level. This aggregation was needed to jointly analyze mode performance and supply but might lead to less precise findings about competing transit service. To increase the accuracy of these comparisons, more granular data is needed. Lastly, future analysis should expand the analysis to a longer panel of observations thereby capturing more variation in weather and other seasonal factors that determine demand for mobility.

## AUTHORS CONTRIBUTION

All authors contributed to all aspects of the study from model development, data processing, to analysis and interpretation of results, and manuscript preparation. All authors reviewed the results and approved the submission of the manuscript.

## ACKNOWLEDGEMENTS
We would like to thank the reviewers for their thoughtful comments that contributed towards improving our manuscript. Amanda Stathopoulos was supported in part by the US National Science Foundation Career Award No. 1847537.

**FIGURES LEGEND**

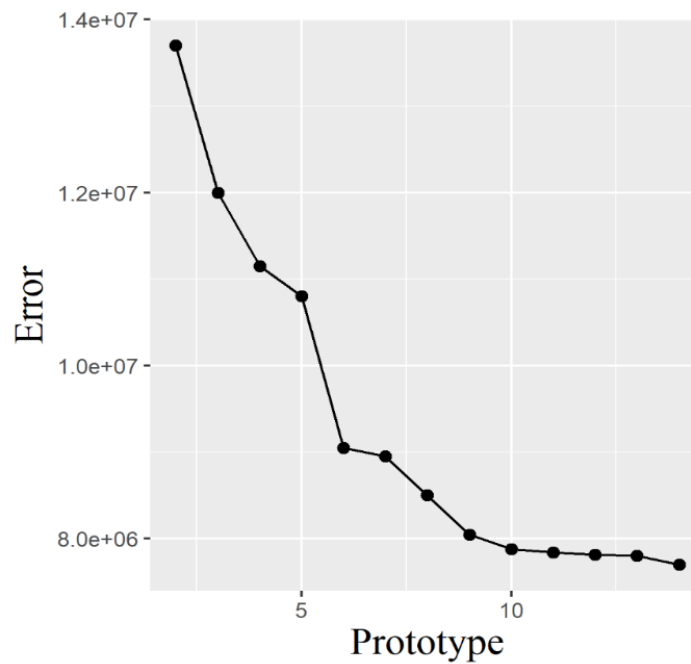

**FIGURE 1 Selection K number of Prototypes**

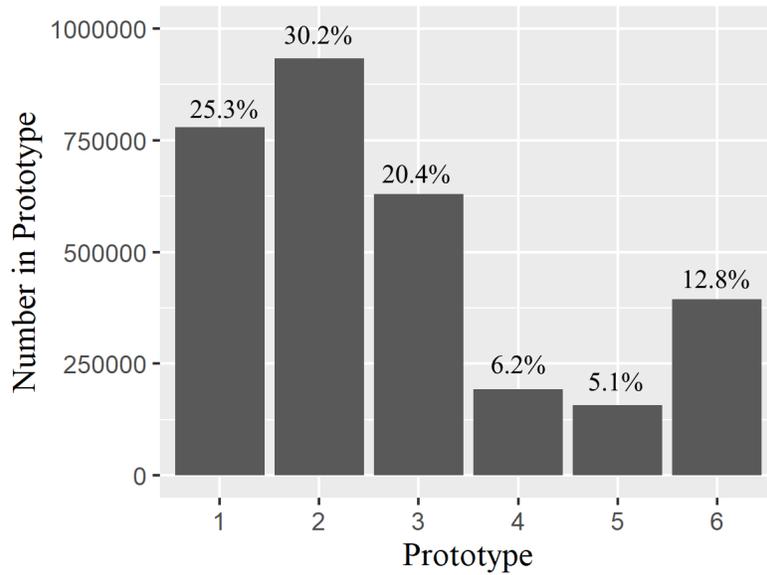

**FIGURE 2 Prototype Shares among total ridesourcing trips**

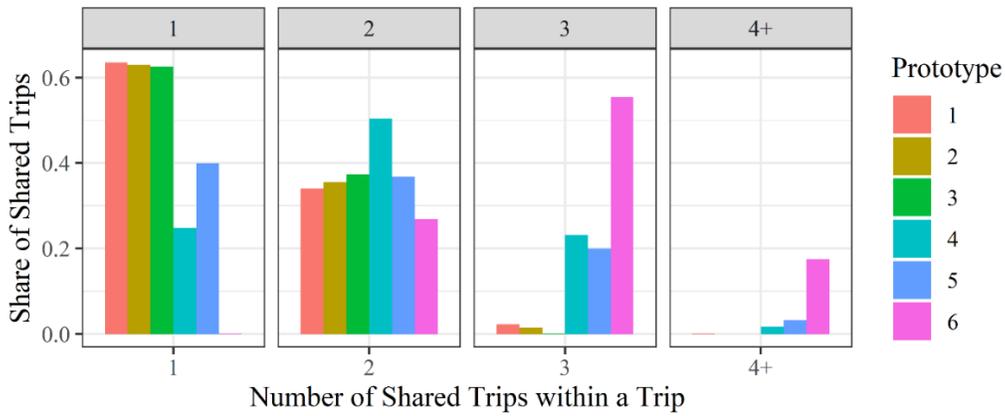

**FIGURE 3 Number of Travelers pooling a ride for Actual Shared Trips**

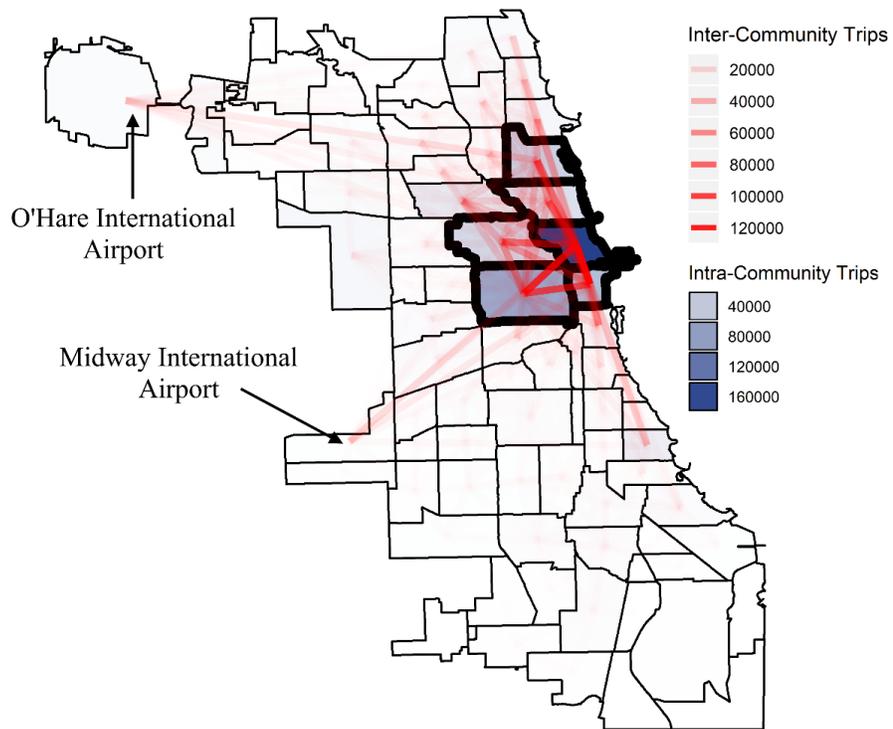

O'Hare International
Airport

Midway International
Airport

**FIGURE 4 Ridesourcing Flows in the City of Chicago with bolded boundaries of prominent Community Areas**